\documentclass[aps,prl,reprint]{revtex4-2}
\usepackage{blindtext}
\usepackage{graphicx}
\usepackage{amssymb}
\usepackage{amsfonts}
\usepackage{amssymb,amsmath}
\usepackage{blindtext}
\usepackage{url}
\usepackage{ mathrsfs }
\usepackage{mathtools}
\usepackage{gensymb}
\usepackage{ textcomp }
\usepackage{hyperref}
\hypersetup{colorlinks=true,urlcolor=blue}

\newcommand{\bs}{\boldsymbol}

\begin{document}

\title{Spatiotemporal patterning of extensile active stresses in microtubule-based active fluids}

\author{Linnea M. Lemma\textsuperscript{1,2}, Minu Varghese\textsuperscript{1}, Tyler D. Ross\textsuperscript{3,4}, Matt Thomson\textsuperscript{4}, Aparna Baskaran\textsuperscript{1},  and Zvonimir Dogic\textsuperscript{2,1}}
\email{zdogic@physics.ucsb.edu}
\affiliation{\textsuperscript{1}Department of Physics, Brandeis University, \textsuperscript{2} Department of Physics, University of California at Santa Barbara, \textsuperscript{3} Department of Computing and Mathematical Sciences, California Institute of Technology, \textsuperscript{4} Division of Biology and Biological Engineering, California Institute of Technology}
\date{September 10, 2022}

\begin{abstract}
Active stresses, which are collectively generated by the motion of energy-consuming rod-like constituents, generate chaotic autonomous flows. Controlling active stresses in space and time is an essential prerequisite for controlling the intrinsically chaotic dynamics of extensile active fluids. We design single-headed kinesin molecular motors that exhibit optically enhanced clustering, and thus enable precise and repeatable spatial and temporal control of extensile active stresses. Such motors enable rapid, reversible switching between flowing and quiescent states. In turn, spatio-temporal patterning of the active stress controls the evolution of the ubiquitous bend-instability of extensile active fluids and determines its critical length dependence. Combining optically controlled clusters with conventional kinesin motors enables one-time switching from contractile to extensile active stresses. These results open a path towards real-time control of the autonomous flows generated by active fluids. 
\end{abstract}

\maketitle

\section{Introduction}
The shapes of cells, tissues, and organs are determined by the patterns of the mesoscopic active stresses, which are collectively generated by microscopic molecular motors \cite{saadaoui2020tensile,martin2010integration,behrndt2012forces,etournay2015interplay,mitchell2022visceral,nerurkar2019molecular}. Developing analogous synthetic force-generating materials is of fundamental interest and is essential for diverse applications~\cite{NeedlemanDogicReview}, ranging from microfluidics and adaptive optics to soft robotics. Active matter, which is an assemblage of microscopic force-generating constituents, provides a promising experimental platform for realizing these goals~\cite{MarchettiReview}. Bulk active materials exhibit diverse dynamical states~\cite{Nedelec1997,schaller2010polar,Sanchez2012,PalacciScience2013, PoonNatComm2018,SoniNatPhys2019}, which can exhibit chaotic or turbulent-like flows. However, developing soft active matter systems is only the first step towards creating life-like synthetic materials; one also needs to harness their force-generating capabilities to stabilize a targeted dynamical state. In this vein, previous work demonstrated that immutable geometrical confinement can stabilize spinning vortices or long-ranged coherent flows~\cite{kuntawu2017Science,GuillamatPNAS2016,WiolandPRL2013,nishiguchi2018engineering}. 

The next challenge is to develop protocols that control transitions between distinct dynamical states {\it in situ}. For inspiration, one can look towards biological cells, which can sense their environment and adjust force patterns accordingly \cite{discher2005tissue}. A first step for achieving analogous capabilities in synthetic materials is to develop active matter whose dynamics are responsive to external cues. Light patterns have been used to control the motility induced phase separation and the structure and dynamics of bacterial ~\cite{PalacciScience2013, PoonNatComm2018,VizsnyiczaiNatComm2017, FrangipaneElife2018,ArltNatComm2018,aubret2018targeted}. Recent advances also demonstrated the optogenetic control of cytoskeletal active matter, including the control of local contractility and aster formation, as well as influencing the motion of topological defects~\cite{Tyler2019,ZhangNatMat2021,qu2021programming}. Here we demonstrate robust and repeatable spatial and temporal control of active stresses in three-dimensional microtubule-based active fluid.To achieve precise control, we develop single-headed kinesin motors that exhibit optically induced enhanced clustering. The unique features of this system enable temporal control of the ubiquitous bend instability of active fluids, and reveal how this instability depends on the system size.   

\subsection{Optical control of kinesin clustering}
Conventional microtubule-based active fluids are powered by clusters of kinesin-1 molecular motors~\cite{Nedelec1997}. Motors within a single cluster bind adjacent microtubules. Depending on the microtubules' relative polarity, the motor clusters induce interfilament sliding and bundle extension ~\cite{lemma2020multiscale}. At high microtubule concentrations, extensile bundles form a percolating network that is continuously reconfigured by the repeating cycles of motor driven bundles extending, buckling, fragmenting and re-annealing~\cite{Sanchez2012}. Such network dynamics drive the turbulent-like flows of the background fluid. Coupling energy-efficient kinesin motors with an ATP regeneration can sustain non-equilibrium dynamics for hours or even days~\cite{chandrakar2022engineering}.

The formation of multimotor clusters is essential for generating active stresses. Isolated motors move along separate filaments, but are unable to generate interfilament sliding. The biotin-streptavidin interaction used to assemble conventional clusters is an essentially irreversible non-covalent bond~\cite{SedlakSciAdv2020}. Decreasing intra-cluster bond strength, decreased the efficiency of interfilament sliding, as was revealed by DNA-based clusters, in which the binding strength can be precisely tuned~\cite{AlexPNAS2021}. Below a minimal bond strength, the clusters were unable to generate active stresses. 

Such considerations suggest that controlling cluster assembly provides a route towards control of active stresses. Indeed, a recent advance fused optically responsive dimerizing domains (improved light induced dimers - iLID) to kinesin motors to establish reversible light control of the assembly of the kinesin clusters~\cite{Tyler2019,iLID2015}. In the absence of photoactivation, the binding affinity of the kinesin-iLID to its polypeptide binding partner, which is attached to another kinesin motor, is low: thus the kinesin motors are largely isolated. They consume ATP and walk along microtubules, but do not induce significant interfilament sliding [Fig. \ref{Fig1}(a-b)]. When photoactivated, the iLID domain changes conformation, exposing the binding site and increasing the affinity for motor pairs to dimerize. The newly formed motor clusters can crosslink and slide microtubules [Fig. \ref{Fig1}(a-b)]. Such opto-kinesin clusters can induce photoactivated contraction of microtubule asters and generate programmable fluid flows \cite{Tyler2019}. 

Microtubule-based active matter is conventionally powered by the N-terminal 401 amino-acid fragment of \emph{Drosophila} kinesin-1 motor~\cite{Nedelec1997,Sanchez2012}. Two 401 amino-acid fragments associate to form a two-headed dimeric kinesin motor, which we refer to as K401. These processive motors are continuously bound to a microtubule as they take about one hundred 8-nm ``hand-over-hand'' steps toward the microtubule plus end, before detaching \cite{Thorn2000,yildiz2004kinesin}. The C-terminal of the K401 motors is labelled with biotin. Adding tetravalent streptavidin induces assembly of higher-order clusters containing several K401 dimer motors, which efficiently power active matter dynamics.   

Microtubule-based active matter can also be powered by a monomeric N-terminal 365 amino-acid fragment of \emph{Drosophila} kinesin-1 motor, which we refer to as K365 ~\cite{chandrakar2022engineering,AlexPNAS2021}. Having a shorter neck linker domain, such fragments do not dimerize. K365 motors are not processive; they detach from a microtubule after each step~\cite{Berliner1995, HancockJOCB1998}. Similar to K401, K365 kinesin motors are labelled with biotin and can assemble into higher order clusters with streptavidin. Notably, the K365 motor clusters are still able to bind multiple filaments and induce their relative sliding, generating active stresses. The dynamics of an active fluid driven by clusters of K365 motors exhibit greater temporal stability of flow speeds and structural length scales, when compared to those driven by K401 motors \cite{chandrakar2022engineering}. Clustering can transform non-processive motors into processive ones~\cite{furuta2013measuring}.

\section{Results} 

\subsection{Steady-state dynamics of active fluids powered by opto-kinesin clusters}

\begin{figure*}
    \centering
    \includegraphics[width=14cm]{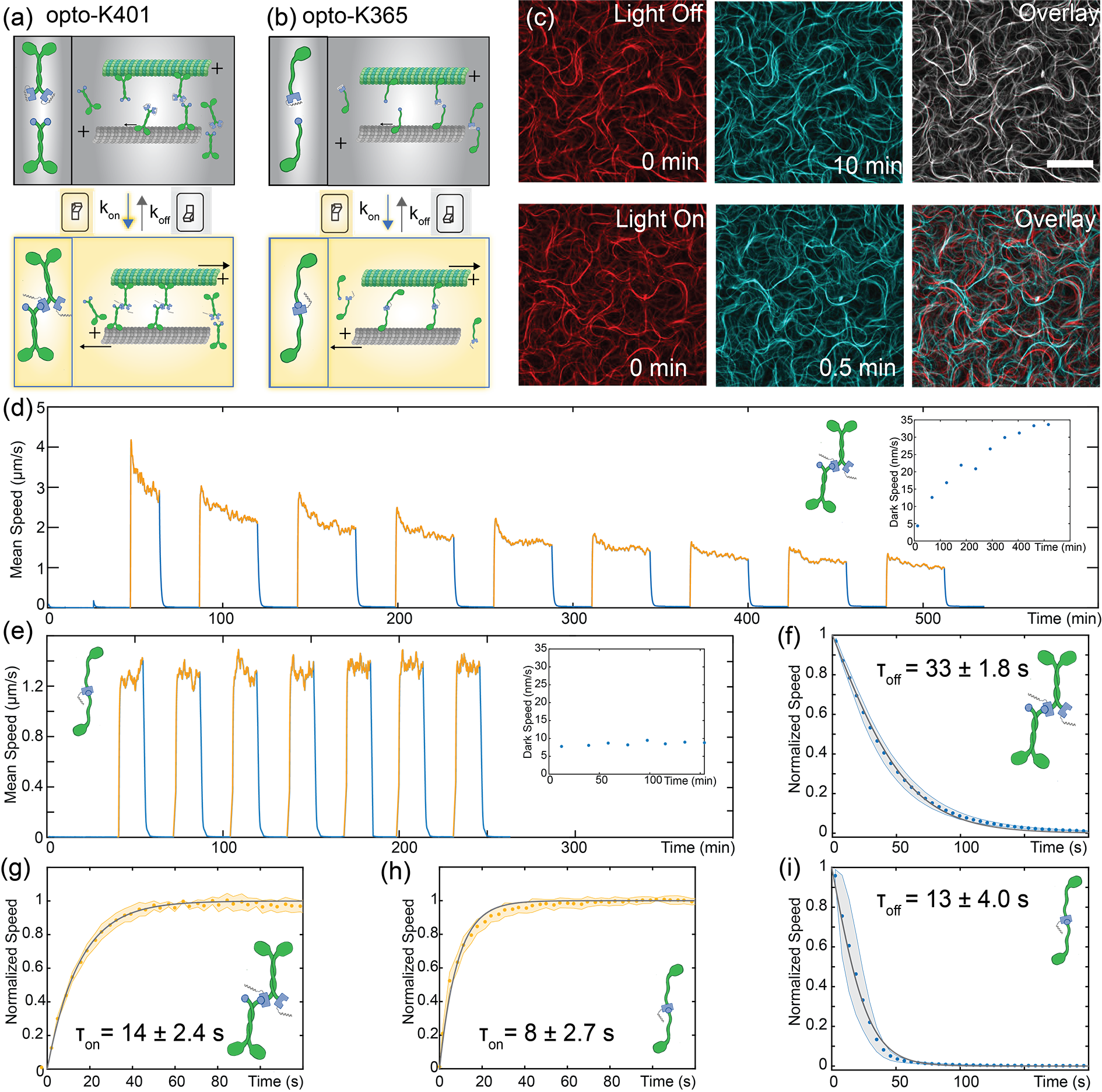}
    \caption{\textbf{Optical switching between active extensile fluid and an inactive solid.} \textbf{(a)} The iLID optogenetic domains are fused to processive kinesin motors. When the protein is illuminated with blue light, the iLID domain changes conformation and the complementary constructs bind to each other. The clustered motors generate interfilament sliding. \textbf{(b)} The iLID domains are fused to a K365 truncated kinesin motor. The K365 motor clusters also dimerize and induce interfilament sliding when exposed to blue light. \textbf{(c)} An active network of  labeled microtubules powered by optogenetic kinesin clusters. Overlaid initial (red) and final frame (cyan) show the network dynamics (red + cyan = gray). top: In the dark, the microtubule bundles do not move. bottom: When exposed to blue light, the bundles extend and buckle generating active flows. Scale bar, 25 $\mu$m. \textbf{(d)} Speed of active flows generated by processive opto-kinesin clusters. Activation is cycled off (blue) and on (gold). In the on-state the light is pulsed for 10 ms every 5 s. Inset: the average network speed in the dark state with subsequent photo-activation cycles. \textbf{(e)} Active flow speeds with K365 opto-kinesin. Inset: dark speed over time. \textbf{(f)} The decay of the flow speed upon turning the light off for processive motors. The shaded region is the standard deviation (n=9). The gray line is a fit to Eq. \ref{vel_off}. \textbf{(g)} For processive motors, the increase in speed upon photoactivation. The gray line is a fit to Eq. \ref{vel_on}. \textbf{(h)} The increase in speed after photoactivation for K365 (n=7).   \textbf{(i)} The decay in the flow speed after deactivation for K365 motors.  }
    \label{Fig1}
\end{figure*}

We first explored the ability of K401-opto clusters  to control extensile dynamics of microtubule-based active fluids [Fig.~\ref{Fig1}(a)]. In presence of blue light opto-clusters generated extending bundles that continuously buckled and annealed [Video 1, Materials and Methods]. Images taken 30 s apart revealed a continuously reconfiguring microtubule network [Fig.~\ref{Fig1}(c) bottom]. Upon deactivation, the microtubule bundles quickly stopped extending and remained frozen in their paused configuration. Images taken 10 min apart demonstrate the lack of measurable dynamics in the absence of optical signal [Fig.~\ref{Fig1}(c) top]. When exposed to continuous blue light, the photoactivated flows decayed rapidly, perhaps due to irreversible crosslinking of motors clusters. Pulsed 10 ms photoactivation every 5 s avoided this issue. Additionally, we optimized the motor concentration to achieve maximal difference between the dark and photoactivated state [Materials and Methods]. 

We measured the speed of the microtubule-powered flows through multiple cycles of photo-activation, using particle image velocimetry (PIV) of the fluorescent microtubule images \cite{thielicke2014pivlab} [Fig.~\ref{Fig1}(d), Fig.~S1, Materials and Methods]. After formation of the bundled network, the sample was alternately exposed to dark and light cycles each lasting 15 minutes and 1 hour respectively. When illuminated active flows increased 100-fold from $\sim20$ nm/s in the dark to $\sim2$ $\mu$m/s. The difference in speeds persisted over several light cycles that spanned the active network lifetime. 

Quantifying the flow speeds, however, revealed subtle variations of photoactivated dynamics over time. First, within a single cycle, the maximum speed was reached within 30 s after photoactivation and then decayed to a plateau. The decay timescale increased with subsequent photo-activation cycles. Second, the plateau speed decreased with each successive photo-activation cycle. Such decay is reminiscent of aging observed in active fluids driven by the conventional K401 motor clusters \cite{chandrakar2022engineering}. Finally, the background speed in the dark state increased throughout the sample lifetime [Fig.~\ref{Fig1}]. The combination of increased dark flows and decreased photoactivated flows reduced the contrast between the dark and the illuminated states [Fig.~S2]. These behaviors present a challenge for applications that require repeatable control of the active stress.

Motivated by the observation that K365 kinesin clusters generate more regular temporal dynamics \cite{chandrakar2022engineering}, we created clusters which fused a monomeric kinesin fragment (K365) to both iLID domains [Fig.~\ref{Fig1}(b), Fig.~S3]. The K365 opto-clusters also powered the extensile dynamics upon photo-activation [Video~2]. After formation of the bundled network, the active fluid was exposed to repeating cycles of photo-activation. In contrast to the K401-opto motors, the K365-opto constructs exhibited a regular, reversible and reproducible response. Photoactivated flows immediately plateaued, at 1.2 $\mu$m/s, while dark state flows were measured $\sim 7$ nm/s. Subsequent photo-activation cycles generated the same speed, with no significant change over the sample lifetime. Finally, the dark state flow speeds did not increase over time, and were effectively zero when compared to enzymatically dead samples [Fig.~\ref{Fig1}(f), Fig.~S4]. Thus, K365 opto-kinesin clusters allow for robust and repeatable spatio-temporal control of active stresses. 

The steady state dynamics of extensile fluids is uniquely suited to characterize the temporal response of clusters to optical stimulus, which is related to the kinetics of the kinesin cluster assembly and disassembly. We measured the timescale for which the flows reached their maximum speed after photo-activation. For both K401-opto and K365-opto motor clusters, the increase in speed was fitted to a bounded exponential function
\begin{equation}
\langle|\bs{v}(t)|\rangle/v_\textrm{max}=1-e^{-t/\tau_{\textrm{on}}}.\label{vel_on}
\end{equation} 
For K401 clusters, $\tau_{\textrm{on}}=14\pm2.4$ s and for K365 clusters $\tau_{\textrm{on}}=8\pm2.7$ s [Fig. \ref{Fig1}(g)(h)]. Similarly, we measured the time scales for the flows to cease after deactivation. We fitted the velocity decays to a sigmoid function: \begin{equation}\langle|\bs{v}(t)|\rangle = \frac{2}{1+e^{-t/\tau_{\textrm{off}}}} ,\label{vel_off}\end{equation}
from which we extracted $\tau_{\textrm{off}}=33\pm1.8$ s for K401 clusters and $\tau_{\textrm{off}}=13\pm4.0$ s for K365 clusters [Fig.~\ref{Fig1}(f)(i)].

\subsection{Controlling the speed of extensile flows}
\begin{figure*}
    \centering
    \includegraphics[width=8cm]{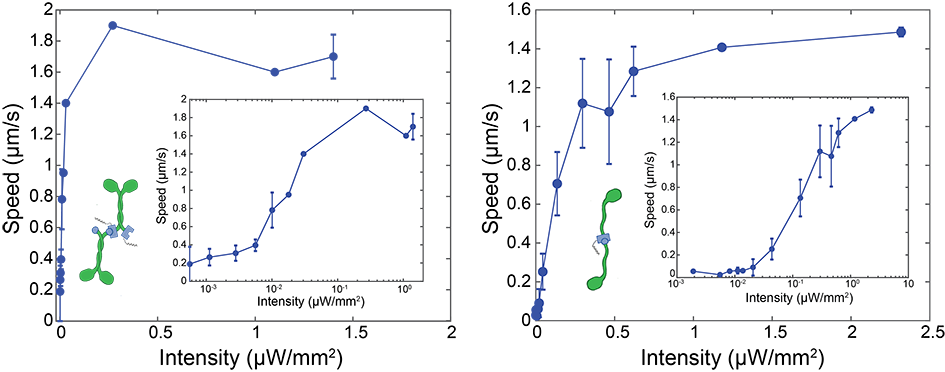}
     \caption{\textbf{Signal intensity controls speed of active flows. } \textbf{(a)} Autonomous flow speed as a function of the light intensity for processive motors (n = 2, 2, 2, 2, 2, 1, 1, 1, 4, 1 Intensity = 0 $\rightarrow$ 2 $\mu$W/mm$^2$). \textbf{(b)} Speed of autonomous flows measured through PIV versus the light intensity for K365 motors (n = 3, 2, 4, 3, 1, 2, 3, 4, 2, 3, 2, 2, 2, Intensity = 0 $\rightarrow$ 2.5 $\mu$W/mm$^2$). Insets: Log-log scaling of axes. Error bars represent standard error on multiple samples' flows.}
    \label{Fig2}
\end{figure*}

We next aimed to control the speed of active flows. Absorbance assays off AsLOV proteins, the native protein from which iLID was derived, indicated that the light intensity controls the fraction of dimerized domains~\cite{Salomon2000}. Thus, we expect that the light intensity also controls the concentration of bound motor clusters, which in turn should control the magnitude of the active stress and the flow speeds. We measured the dependence of the autonomous flow speeds on the intensity of 488 nm illuminating light at the sample plane [Fig.~\ref{Fig2}, SI: Experimental Methods]. With increasing light intensity, the flow speed increased monotonically until saturating at 0.4 $\mu$W/mm$^2$ for K401-opto clusters and 1 $\mu$W/mm$^2$ for K365-opto clusters [Fig.~\ref{Fig2}(a-b)]. Below 0.01 $\mu$W/mm$^2$ the measured flows asymptotically approached the dark state [Fig.~\ref{Fig2}(a)(b) inset]. 

\subsection{Controlling the extensile bend instability}
\begin{figure*}
    \centering
    \includegraphics[width=16cm]{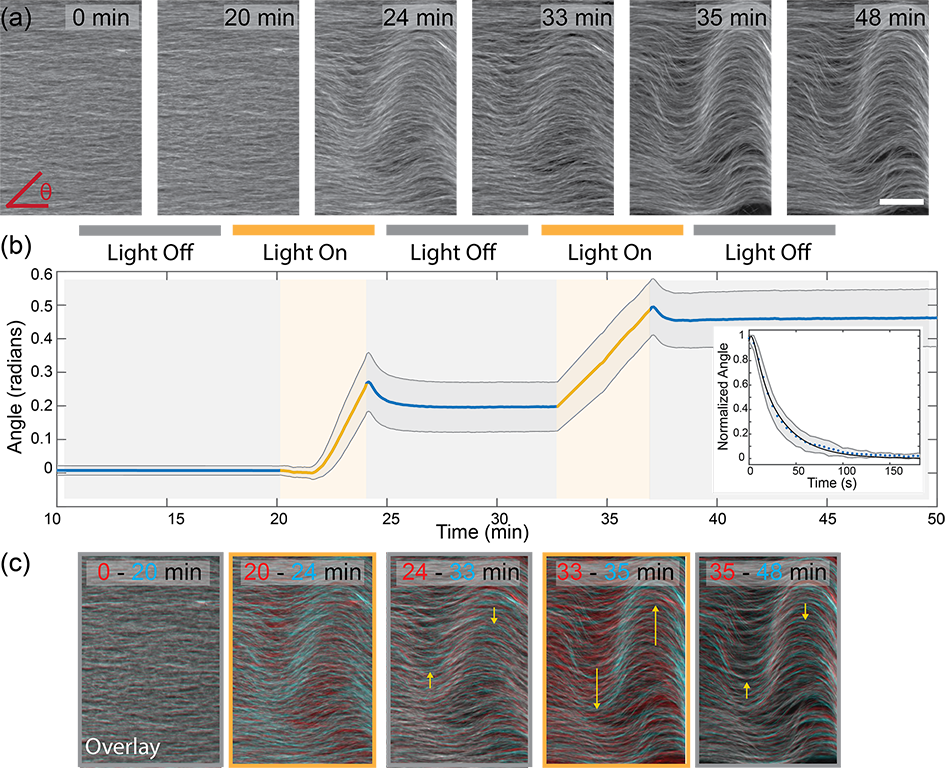}
     \caption{\textbf{Controlling the bend instability } \textbf{(a)} Time series of a flow aligned opto-kinesin microtubule network exposed to alternating dark and light conditions. Scale bar, 200 $\mu$m. \textbf{(b)} The time-evolution of the average angle for two photo-activation cycles. Inset: The relaxation dynamics of the bend instability after deactivation. The average angle as a function of time elapsed after deactivation. The shaded region is standard deviation (n=5). \textbf{(c)} Overlays of time points in (a) with early time in red shows relative microtubule movement across each light cycle. Yellow arrows direct the eye forward in time. }
    \label{InstabilityFig}
\end{figure*}

To demonstrate temporal control of active stresses, we used K365 opto-clusters to probe the bend instability of shear-aligned microtubules, which is a ubiquitous feature of extensile active fluids~\cite{SimhaPRL2002,SokolovPRX2019,martinez2019selection,Pooja2020}. Once aligned due to shear flow, microtubules remained quiescent in the absence of light [Fig.\ref{InstabilityFig}(a)]. When exposed to blue light, the opto-kinesin clusters generated extensile sliding, which in turn powered the growth of the bend instability. Upon deactivation, the bend instability stopped [Fig.~\ref{InstabilityFig}(b), Video 3]. The individual microtubule bundles within the deactivated network straightened slightly, but the material only partially relaxed towards its initial uniformly aligned state. Similar deformation dynamics occurred over multiple cycles of photo-activation, with motor clusters pushing the microtubules into an active bend, and the network partially relaxing when deactivated. 

To quantify the bend instability, we measured the average filament angle $\langle\theta\rangle$ away from the initial alignment, by calculating the local structure tensor from fluorescence images [Fig.~\ref{InstabilityFig}(b), Fig.~S5]. During the photo-activation, the average angle $\langle\theta\rangle$ grew as the bend instability developed. Upon deactivation, the average angle decayed, but after several minutes attained a finite-value plateau. Such dynamics is seen by overlaying the initial and final images upon photoactivation [Fig.~\ref{InstabilityFig}(c)]

The decay in the average angle upon deactivation was described by:
\begin{equation}
    \langle\theta(t)\rangle = M(t)A(t)+R(t).
\end{equation}
Here, $M(t)$ describes the unbinding kinetics of the motor clusters, $A(t)$ describes the activity-driven angle growth and $R(t)$ describes the relaxation of the passive microtubule network.
The activity drives linear growth such that $A(t) = \dot{\gamma} t$, where $\dot{\gamma} = 0.002$ rad/s is the measured angular growth rate [Fig. S6]. Upon deactivation, growth slowed due to unbinding of opto-clusters. To account for this effect, we multiply the growth rate by the rate at which the clusters fall apart. Studies of unbinding kinetics of the light activated domain revealed exponential decay, $M(t)=e^{-t/\tau}$ where $\tau$ = 24 s [Fig. S7, SI.1] ~\cite{HallettACSSynthBio2016}. Modeling the microtubule network as an elastic solid, predicts $R(t)=e^{-t/r}$ where $r$ is the elastic relaxation time [SI]. Thus:
\begin{equation}
    \langle\theta\rangle=e^{-t/\tau}\dot{\gamma} t + C e^{-t/r}
    \label{elastic_model}
\end{equation}
where $\tau$ is the characteristic unbinding time for the motors and $C$ is a scaling fit parameter. To fit the data, we fixed $\dot{\gamma}=0.002$ rad/s, and left the time scales $\tau$ and $r$ as free parameters [Fig.~\ref{Fig3}(b) inset, Fig.~S8]. Experimental fits yielded $\tau=17\pm5.5$ s which is within error of the previously reported unbinding kinetics of the iLID domain \cite{HallettACSSynthBio2016}. We also extracted, the elastic relaxation time $r = 17\pm 9.8$ s. 

\subsection{Spatial patterning of active stresses}
Theory predicts that the bend instability only develops when the sample size is larger than a critical length scale, which in turn is determined by the ratio of the sample elasticity to the activity~\cite{SimhaPRL2002,GaoPRL2015}. Opto-kinesin clusters enable application of spatially patterned stress, which can elucidate how the interplay between geometry and activity controls the onset of the bend instability. 

\begin{figure*}
    \centering
    \includegraphics[width=16cm]{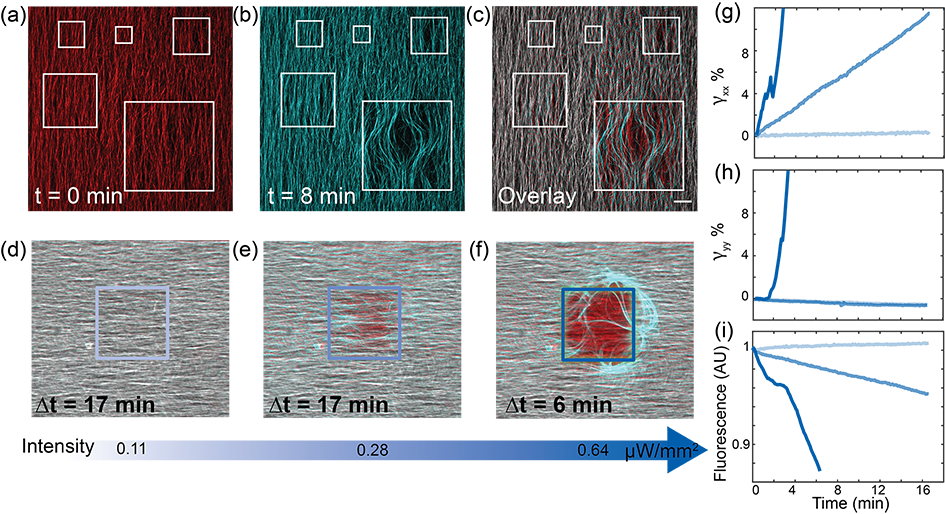}
    \caption{\textbf{Instability of confined active fluid}
    (a)  White squares indicate activation regions from 50 $\mu$m$^2$ to 500 $\mu$m$^2$ of aligned microtubules. (b) Microtubule network after 8 min of constant photoactivation within the confined regions. (c) The overlay of the initial and final images. (d-f) The dynamics of the quiescent, sliding and buckling regime is illustrated by overlaying of initial (red) and final (cyan) state of a flow aligned network. 400 $\times$ 400 $\mu$m square region was activated. Gray indicates no movement of filaments (red + cyan = white). (g)  $\gamma_{xx}$ within the activated region over time for each activation intensity. (h) $\gamma_{yy}$ within the activated region over time for the three regimes. (i) The average fluorescence intensity inside the activated region over time for the three regimes}

    \label{Fig3}
\end{figure*}

Shear flow associated with chamber loading induced initial alignment of microtubule bundles. When uniformly illuminated, such samples  exhibited the previously-described bend instability [Fig.~\ref{Fig3}]~\cite{Pooja2020}. We  used a laser scanning confocal microscope to photoactivate square regions ranging in size from $50 \times 50$ $\mu$m$^2$ to $500 \times 500$ $\mu$m$^2$. At a single photoactivation intensity, the $500 \times 500$ $\mu$m$^2$ region exhibited the bend instability, while the smaller regions remained quiescent, demonstrating a size-dependent instability [Fig.~\ref{Fig3}(a-c), Video~4].

To quantify the size-dependent instability, we systematically changed the photoactivation intensity within a single region. We placed an opaque mask above the sample with a $400 \times 400$ $\mu$m$^2$ square opening. Exposing such samples to light photoactivated the exposed area, while leaving the rest of the sample inactive. With sequentially increasing light intensity, we observed three regimes [Video~5]. At very low intensities, the microtubules inside the activated region remained stationary [Fig.~\ref{Fig3}(d)]. Beyond a threshold intensity, the microtubules started sliding past each other along the alignment direction, but there was no deformation in the perpendicular direction [Fig.~\ref{Fig3}(e)]. Finally, at high intensities, the microtubules buckled in the direction perpendicular to alignment [Fig.~\ref{Fig3}(f)]. We refer to these as the quiescent, sliding, and buckling regimes. 

To quantify these observations, we measured the microtubule displacement field after the region was photoactivated for a defined time [Fig. S9]. The direction of the initial alignment was the $x$-axis, while $y$-axis was the perpendicular direction in the image plane. For the purposes of this analysis, we neglected the out of plane component. We calculated the build up of strain over time by averaging $\gamma_{xx}$ and $\gamma_{yy}$ over the photoactivated area [Fig.~\ref{Fig3}(g-h)]. In the quiescent regime $\gamma_{xx}$ remained zero, indicating lack of any measurable motor-driven dynamics. Interestingly, $\gamma_{yy}$ decreased over time [Fig.~\ref{Fig3}(h)]. This can be attributed to the depletion induced contraction~\cite{braun2016entropic}. Indeed, a flow-aligned sample that was never photoactivated exhibited a slight contraction, pulling toward the chamber center [Fig.~S10]. In the sliding regime, $\gamma_{xx}$ increased linearly with time, indicating a constant extension rate. Concurrently, $\gamma_{yy}$ decreased to the same extent as the quiescent regime, indicating lack of motor-driven transverse dynamics. In the buckling regime, $\gamma_{xx}$ grew quickly, increasing by more than 100\% in 5 min. Simultaneously, $\gamma_{yy}$ became positive and grew rapidly, indicating filament motion perpendicular to the direction of extension.

We estimated the material flux through the boundary of the photoactivated region, by measuring the change in the average fluorescent intensity of labeled microtubules within the exposed region [Fig.~\ref{Fig3} (i)]. Similar to the strain analysis results, the fluorescent intensity in the quiescent regime showed a slight increase of material over time due to the depletion induced contraction. In the sliding regime, the intensity decreased linearly as microtubules extended out of the photoactivated region along the $x$-axis. In the buckling regime, the intensity decreased rapidly as the bend instability pushed the microtubules out of the photo-activation region. When expelled in the background, microtubules became stationary and did not return to the exposed region. Thus, with continued illumination, the photo-activated region eventually became devoid of microtubules [Video~6].

\begin{figure*}
    \centering
    \includegraphics[width=8cm]{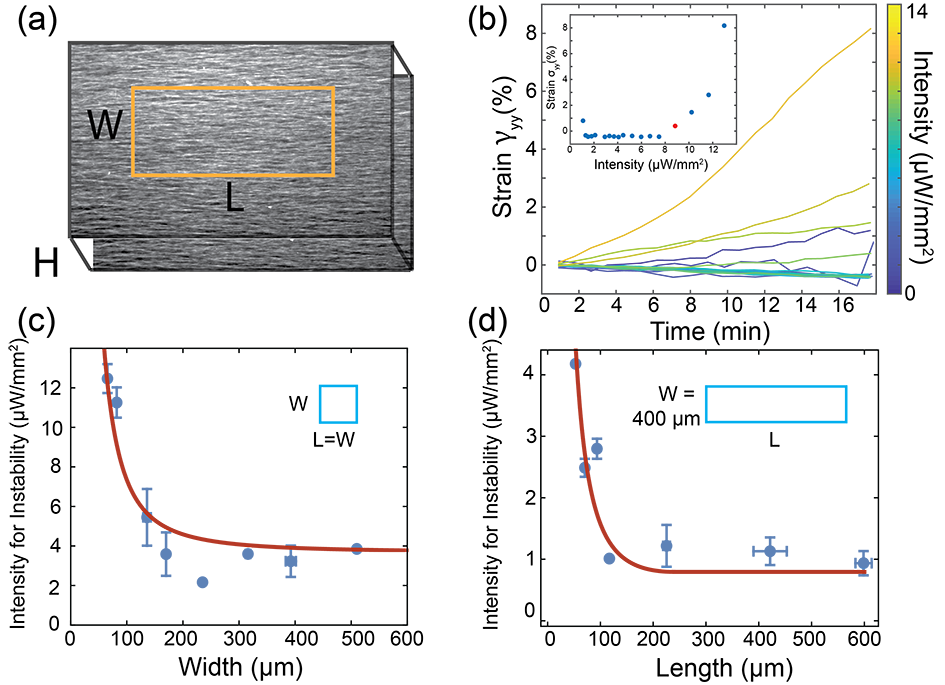}
     \caption{\textbf{Length scale of the bend instability} \textbf{(a)} A rectangular region (length $L$ and width $W$) of flow-aligned microtubules is activated. The system confinement is three-dimensional, with the height $H$ defined as the height of the chamber. \textbf{(b)} $\gamma_{yy}$ plotted over time with increasing intensity (color). Inset: the strain after 17 min of photoactivation versus intensity. The bend instability is defined at the lowest intensity for which $\gamma_{yy}$ increases by more than 0.5\% over 17 min, indicated by the red dot. \textbf{(c)} The threshold intensity for instability plotted versus the $L=W$ of a square activated region in a chamber $H=100$ $\mu$m. Red line is fit to theory Eq. \ref{Minu1} with $a=1.1~$nW$^{-1}$. Error bars indicate standard error on multiple activated areas (n = 2, 2, 4, 3, 1, 2, 3, 1 from W = 50 $\mu$m - 500 $\mu$m). \textbf{(d)} The threshold intensity for instability for $W=400$ $\mu$m plotted versus the length $L$ of the activated region in a chamber $H=300$ $\mu$m. Error bars indicate standard erro on multiple activated areas (n = 1, 2, 2, 1, 3, 2, 3 from L = 50 $\mu$m $\rightarrow$ 600 $\mu$m). Red line is fit to theory Eq. \ref{Minu1} with $a=0.93~$nW$^{-1}$. }
    \label{Fig4}
\end{figure*}

The combination of the width $W$ and the height $H$ of the microfluidic channel determines the length scale of the bend instability~\cite{Pooja2020}. We illuminated aligned samples with rectangular patterns, with length $L$ along the alignment direction and width $W$ perpendicular to the alignment [Fig.~\ref{Fig4}(a)]. We systematically increased the light intensity to find the threshold activity required for the bend instability. The signature of bending is the growth of $\gamma_{yy}$. We defined a threshold intensity for the bend instability to be when the strain $\gamma_{yy}$ exceeded 0.5\% over a 17 minute activation [Fig.~\ref{Fig4} (b), Fig.~S11]. The intensity of light required to generate buckling in this time depended on the size of the illuminated region. For example, the intensity required to generate buckling of $L\sim 100$ $\mu$m region was three times larger than for $L\sim500$ $\mu$m squares [Fig. \ref{Fig4}(c)]. 

The critical length scale required for instability can be calculated using a hydrodynamic model [SI: Hydrodynamic Model]. The hydrodynamic theory predicts that the instability is controlled by a dimensionless activity parameter $\alpha_{\textrm{eff}} = (\alpha(S_0+\xi))/(2\eta D_R \kappa)$. For an aligned nematic channel of dimensions $L$, $W$ and $H$, the threshold for the onset of the instability is given by: 
\begin{widetext}
\begin{equation}
    \frac{I_{\textrm{threshold}}}{a}=\alpha_{\textrm{eff}}=\left\{
   \begin{array}{ll}
    \pi^2L^2\left(\frac{1}{L^2}+\frac{1}{W^2}+\frac{1}{H^2}\right)^2, & \hspace{1 cm}  \frac{1}{L^2}>\frac{1}{H^2}+\frac{1}{W^2}\\
    4\pi^2\left(\frac{1}{W^2}+\frac{1}{H^2}\right) , & \hspace{1 cm}  \frac{1}{L^2}<\frac{1}{H^2}+\frac{1}{W^2}
    \end{array}
    \right\}
    \label{Minu1}
\end{equation}
\end{widetext}
where, $S_0$ is the magnitude of initial order, $\xi$ is the flow alignment parameter, $D_R$ is the rotational diffusion, $\kappa$ is the elastic constant, $\eta$ is the viscosity of the active fluid and $L$, $W$ and $H$ are the dimensions of the activated region [Fig.~\ref{Fig4}(a)]. 

If we assume that the active stress is linearly proportional to the intensity of the optical signal, $I= a \alpha_\textrm{eff}$, where $a$ is a fit parameter, the hydrodynamic model explains the size dependence of the bend instability [Fig. \ref{Fig4}(c)]. The theory further predicts that if the confinement perpendicular to the direction of alignment is held constant $W=400$ $\mu$m while the length $L$ of the illuminated region varies, the threshold intensity will saturate when
$1/L^2 = 1/W^2+1/H^2$ to an effective activity $4\pi^2\left(1/W^2+1/H^2\right)$. Theoretical predictions agree with the experiments, where $a = 1.1$ nW and $a=0.93$ nW for the respective fits  [Eq.~\ref{Minu1}, Fig.~\ref{Fig4}(c)(d)].  

\subsection{Tuning the stress from contractile to extensile}
\begin{figure*}
    \centering
    \includegraphics[width=16cm]{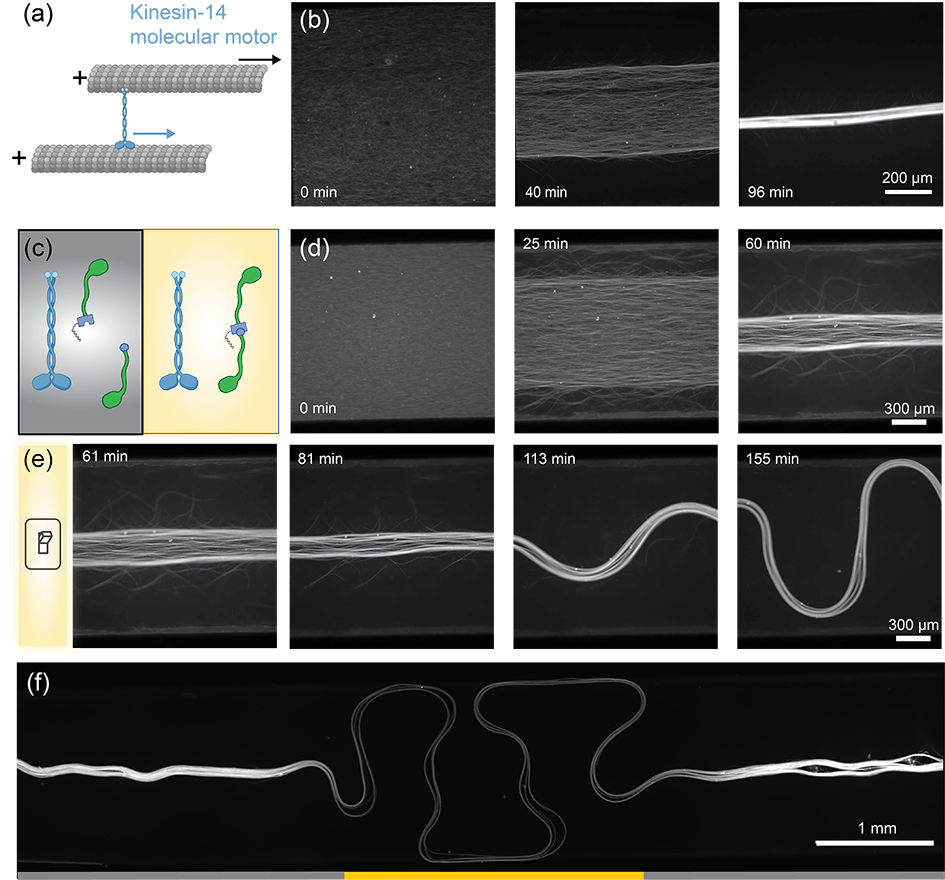}
  \caption{\textbf{Optogenetic switch from contractile to extensile active stress.} \textbf{(a)} Kinesin-14 motor has a passive microtubule binding domain and a motor domain that steps toward the microtubule minus end  (blue arrow). The kinesin drives interfilament sliding (black arrow). \textbf{(b)} A time series of the kinesin-14-driven microtubule contractile dynamics of an aligned microtubule gel. \textbf{(c)} Microtubule-based active matter powered by kinesin-14 and K365 opto-kinesin clusters. In the dark state, kinesin-14 dominates system dynamics, generating contraction. In the illuminated state, opto-kinesin clusters generate competing extensile stresses. \textbf{(d)} A two motor-network with deactivated opto-kinesin clusters contracts. \textbf{(e)} The contracting network is locally photoactivated with 10 $\mu$W/mm$^2$ at 61 min, and the opto-kinesin induce extension. A time series shows the transition to extensile, buckling behavior. \textbf{(f)} Entire sample after photo-activation at t = 156 min. The yellow region has been illuminated. Outside this region, opto-kinesin remain deactivated.}
    \label{Fig5}
\end{figure*}

So far, we demonstrated quantitative control of the extensile active stresses in space and time. In comparison, biological cells are able to control not only the magnitude but also the sign of the active stress to perform complex functions. Although both contraction and extension have been observed in a wide variety of cytoskeletal active matter systems \cite{Peter-eLife2015, KumarSciAdv2018, MizunoScience2007}, an on-demand switching between these two regimes has yet to be realized. Motivated by such reasoning, we explored the dynamics of a multi-motor material composed of K365 opto-kinesin clusters and full length kinesin-14. 
The latter is a minus-end directed molecular dimer motor with a passive microtubule-binding domain, a long neck linker, and a motor domain that steps processively at $\sim20$ nm/s  [Fig.~\ref{Fig5}(a)] \cite{HentrichJOCB2010,claireJOCB1997}. Thus, kinesin-14 can simultaneously passively bind one microtubule and advance towards the minus end of another microtubule, inducing their relative microtubule sliding \cite{CaiMBOC2009}. Kinesin-14 powers interfilament sliding in aligned microtubule networks that is independent of the local filament polarity \cite{BezSebastian}. 

In the presence of a depletion agent, kinesin-14 alone contracted a microtubule network [Material and Methods]. The active contraction led to the formation of a single macroscopic bundle aligned with the chamber walls, which continued contracting for tens of hours [Fig.~\ref{Fig5}(b), Video~7]. Next, we combined the kinesin-14 contracting network with K365 opto-kinesin [Fig.~\ref{Fig5}(c)]. When left in the dark, the material still formed a contracting network, as it would without the opto-kinesin [Fig.~\ref{Fig5}(d)]. However, upon illumination, the opto-kinesin clusters generated their own active stresses [Fig.~\ref{Fig5}(e), Video~8]. The system exhibited extension and buckling in the photoactivated region which suggested extensile active stresses. The extensile behavior was localized to the activated region, thus the system simultaneously exhibited spatially distinct regions of extensile and contractile active stress [Fig.~\ref{Fig5}(f)]. On-demand transition from contractile to extensile stresses opens the door for new fundamental studies related to the sign of active stress, as well as engineering applications. However, the transition was not reversible, highlighting that the microscopic mechanism requires further study. 

\section{Discussion}
Building on the recent developments, we characterized opto-kinesin clusters of non-processive K365 motors, showing that they are optimized for controlling extensile active stresses. These motors add to a growing toolbox of optical control of active dynamics \cite{ArltNatComm2018, Tyler2019, ZhangNatMat2021}. The K365 opto-clusters induced up to a 200-fold increase in the speed of autonomous flows between a dark and a fully illuminated state. Importantly, gels powered with K365 clusters displayed greater temporal regularity when compared to K401 clusters, a finding that parallels the analogous observation of conventional streptavidin-bound clusters~\cite{chandrakar2022engineering}.

Kinesin clusters can both generate active stresses and passively cross-link microtubule bundles \cite{GagnonPRL2020, sarfati2021crosslinking, NajmaArxiv2021}. For example, with decreasing ATP concentration, the motors' primary role switches from generating active stresses to cross-linking filaments~\cite{GagnonPRL2020}. Consequently, ATP controls a transition between a quiescent elastic solid and a spontaneously flowing active fluid. Opto-kinesin clusters provide a dynamical switch between an active fluid and a quiescent, passive state. In principle, in the dark state, the K365 opto-kinesin have a reduced tendency to dimerize. Thus, one might expect the microtubule network in the dark state to remain fluid-like. However, we found that upon deactivation, the network froze, and the bundles retained their structure. Upon re-activation, the flows were almost perfectly correlated with the flow field measured just before deactivation [Fig. S12]. This behavior suggests, that even in the absence of motor crosslinks, a microtubule network is an elastic solid that is held together by PEG-induced depletion. 

The properties of opto-kinesin provide insight into the bend instability, which is a ubiquitous feature of extensile active fluids~\cite{SimhaPRL2002,martinez2019selection,Pooja2020,SokolovPRX2019}. In conventional systems, a uniformly aligned state is unstable to spontaneously growing bend deformation that increase in time and eventually lead to turbulent-like dynamics. Opto-kinesin clusters enable one to pause the bend generation. Once deactivated, we observed only partial relaxation of the deformed network, wherein the microtubule bundles locally straightened as if released from tension. In comparison, a deformed conventional liquid crystal, relaxes to a uniformly aligned state. The partial relaxation again demonstrates that microtubule-based fluids have complex viscoelastic and plastic properties, that remain largely unexplored.    

 Changing the photoactivation intensity of spatially illuminated shear-aligned sample, revealed a size-dependent lower critical illumination intensity, below which the bend instability was not observed. This is in agreement with theory, which predicts a suppression of the bend instability for active nematics that are confined below a critical length scale~\cite{SimhaPRL2002,GaoPRL2015}. In comparison, the size-dependent suppression of the instability was not experimentally observed for shear aligned microtubule fluids confined in microfluidic channels ~\cite{Pooja2020}. The notable difference between the two systems is the nature of the boundaries. Microfluidic channels aligned the filaments, but compressible isotropic active fluids could easily separate from the wall, leaving behind regions devoid of active material. In comparison, partially illuminated aligned gels remained physically connected to the background elastic network. When a confined region was activated below the instability threshold for long times, the region did eventually buckle [Video 9]. However, in this regime microtubules slid against each other, leading to net outflux of the material, suggesting that the material properties of the network changed, before the onset of the instability. 

Light intensity controlled the speed of the active flows. Two observations suggest a complex relationship between measured flow speeds and activity that warrants further studies. The active fluids exhibited a hysteresis in speeds [Fig.~S13(a)]. Increasing photoactivation intensity from low intensities increased the speed of the fluid flows. However, after saturating the system and returning to low intensities, the fluid speed was five times higher than the original photoactivation at the same intensity. In between photoactivation events, the dark speed remained constant and near zero, indicating that the motors returned to their low affinity dark state. There was a noticeable change in the gel structure between the early and late low-light activation [Fig.~S13(b)]. These data suggest that the material properties of the network, rather than the motors contribute to the hysteresis. As a possible explanation, we suggest that the motors are controlling strain rate, which can lead to different fluid flows depending on the microtubule bundle size and buckling length. Second, we observed that the intensity for the saturation in speed of the bulk fluid flow was lower than the intensities which induced sliding and buckling in the confined system. We attribute this to the hydrodynamic constraints of the system in which the activity (or intensity) required for saturation is inversely proportional to the confinement. 

Finally, we note that the K401 motors drove extensile active fluids for longer times and exhibited faster flows than the K365 counterparts. These features will likely be useful for some applications. In this study, we focused on the K365 motors, which provide reproducible response to photoactivation and therefore precise spatio-temporal control over extensile active stress. 
 

Light intensity controlled the speed of the active flows. Two observations suggest a complex relationship between measured flow speeds and activity that warrants further studies. The active fluids exhibited a hysteresis in speeds [Fig.~S13(a)]. Increasing photoactivation intensity from low intensities increased the speed of the fluid flows. However, after saturating the system and returning to low intensities, the fluid speed was five times higher than the original photoactivation at the same intensity. In between photoactivation events, the dark speed remained constant and near zero, indicating that the motors returned to their low affinity dark state. There was a noticeable change in the gel structure between the early and late low-light activation [Fig.~S13(b)]. These data suggest that the material properties of the network, rather than the motors contribute to the hysteresis. As a possible explanation, we suggest that the motors are controlling strain rate, which can lead to different fluid flows depending on the microtubule bundle size and buckling length. Second, we observed that the intensity for the saturation in speed of the bulk fluid flow was lower than the intensities which induced sliding and buckling in the confined system. We attribute this to the hydrodynamic constraints of the system in which the activity (or intensity) required for saturation is inversely proportional to the confinement. 

Finally, we note that the K401 motors drove extensile active fluids for longer times and exhibited faster flows than the K365 counterparts. These features will likely be useful for some applications. In this study, we focused on the K365 motors, which provide reproducible response to photoactivation and therefore precise spatio-temporal control over extensile active stress. 

\subsection{Acknowledgements}
We acknowledge useful discussions with Seth Fraden. We thank Claire E. Walczak and Stephanie C. Ems-McClung for the gift of kinesin-14 protein. We thank Bezia Lemma for help with kinesin-14 experiments. We acknowledge the use of a MRSEC optical and biosynthesis facility supported by NSF-MRSEC-2011846. We also acknowledge the use of the NRI-MCDB Microscopy Facility and the Resonant Scanning Confocal at UCSB supported by NSF MRI grant 1625770. 

\section{Figures}

\clearpage
\pagebreak
%

\clearpage
\newpage
\onecolumngrid

\setcounter{equation}{0}
\setcounter{figure}{0}
\setcounter{table}{0}
\setcounter{page}{1}

\renewcommand{\theequation}{S\arabic{equation}}
\renewcommand{\thefigure}{S\arabic{figure}}

\begin{center}
\textbf{\large Spatiotemporal patterning of extensile active stresses in microtubule-based active fluids\\Supplementary Material}\\[.2cm]
\end{center}

\begin{center}
Linnea M. Lemma, Minu Varghese, Tyler D. Ross, Matt Thomson, Aparna Baskaran,  and Zvonimir Dogic
\end{center}

\section{Materials and Methods}
\subsection{Protein Preparation}

GMPCPP stabilized microtubules with an average length of 1 $\mu$m were prepared as previously described \cite{tayar2022assembling}.The plasmids encoding for processive opto-kinesin K401 (K401-iLID: Addgene 122484 and K401-micro: Addgene 122485; gifts from Tyler Ross and Matt Thomson) were transformed into BL21 cells. After culturing for 4 hours in 2XYT media, expression was induced at OD 0.6 with IPTG and grown for 16 hours at 18$^{\circ}$C. Cells were lysed using a Branson tip sonicator 550 with a 0.25 inch tip for 10 s at 30\% power with 50 s recovery for 6 cycles. Lysate was then clarified by spinning at 100,000 RCF for 30 minutes. Purification was performed using an AKTA Fast Protein Liquid Chromatography system (GE Healthcare) with a 1 mL Nickel column (HisTrap, GE Healthcare) \cite{tayar2022assembling, Tyler2019}. The wash buffer used was 20 mM imidazole, 50 mM Sodium phosphate, 250 mM NaCl, 4 mM MgCl2, 0.05mM ATP, 5 mM $\beta$mercaptoethanol, 5\% w/v glycerol, pH 7.2. The elution buffer was the same as the wash buffer but with 500 mM imidazole. For the K401-micro construct, the MBP tag was digested with TEV protease and removed by passing over a Nickel column. The buffer was exchanged into wash buffer and then diluted to 50\% v/v glycerol. The kinesin motors were aliquoted, flash frozen in liquid nitrogen and stored at -80$^\circ$C. 

To construct the K365-iLID and K365-micro plasmids (Addgene 188456, 188457), we PCR amplified the desired regions [Fig.~S3] from the K401-iLID plasmid (Addgene 122484, gift from Matt Thomson) and the K401-micro plasmid (Addgene 188457, gift from Matt Thomson) with Q5 High-Fidelity DNA polymerase (New England BioLabs). We then used HiFi DNA assembly kit (New England BioLabs) to circularize the product.  We transformed the plasmids into BL21 cells, and expressed and purified the protein following the same protocol as the processive motors described above \cite{Tyler2019}. 

The active mixture was prepared without kinesin and microtubules at 1.4 mM ATP~\cite{tayar2022assembling}.  Briefly, an active mixture was prepared in M2B (80 mM PIPES, 2 mM MgCl2, 1 mM EGTA, pH 6.8) with a depletion agent to induce microtubule bundling, an ATP regeneration system to maintain activity and oxygen scavenging system to prevent photobleaching: 0.8\% w/v PEG (35 kDa, Sigma Aldrich), 26.6 mM PEP (Beantown Chemicals product \#129745), 1.4 mM ATP (Sigma Aldrich), 6.7 mg/mL glucose, 0.4 mg/mL glucose catalase, 0.08 mg/mL glucose oxidase, 5 mM MgCl$_2$. This active mixture was prepared once for all experiments with opto-kinesin. After mixing, the reagents were flash frozen in liquid nitrogen in 5 $\mu$L aliquots and stored at -80$^{\circ}$C until the day of experiments. 

\subsection{Chamber Construction}
Glass slides were treated with an acrylamide brush to prevent proteins sticking~\cite{LauEPL2009,tayar2022assembling}. The opto-kinesin constructs were especially sensitive to high coverage surface treatments, when  compared to the conventional kinesin-streptavadin clusters. Parafilm was cut into chambers 1, 2 or 3 mm in width and sandwiched between two acrylamide slides. Using a hot plate, the parafilm sandwich was heated to  60$^\circ$C. While on the hot plate, we gently pressed down on the parafilm spacer using the blunt end of an eppendorf 0.5 mL tube until the parafilm turned translucent. For experiments where the height of the chamber was altered, multiple parafilm spacers were carefully stacked on top of each other and the chamber was constructed as above. 

\subsection{Optical Microscopy}
Images were taken on Nikon Ti2-Eclipse inverted microscope equipped with either a Photometrics Prime 95b or an Andor Zyla 5.5 sCMOS camera. Additionally, we used a Leica Resonant SP8 scanning confocal with Stellaris white light laser for some confined activation and subsequent imaging. 

\subsection{Sample Preparation}
All the samples were prepared in conditions where the sample was only exposed only to red light from an red LED-equipped desk lamp. Microtubules, active mixture aliquots and kinesin motors were rapidly thawed. Opto-kinesin motor constructs were added to 300 nM concentration. This was the minimal concentration at which we reliably obtained extensile active gels while photoactivated. Higher concentrations of motors increased the dark state velocities and decreased the sample lifetime. Components were mixed for final desired concentrations and immediately loaded into chamber. The chamber was sealed using a fast setting silicone polymer (Picodent Twinsil Speed). 

\subsection{Sample Preparation of Multimotor Composite}
The multimotor composite was prepared in an active mixture at 80 mM PIPES, 5 mM magnesium chloride, 1 mM EGTA, 0.034\% pyruvate kinase, 52 mM PEP, 1.4 mM ATP, 0.5 mg/mL tubulin and 0.1\% PEG. An oxygen scavenging system was also added to prevent photobleaching (6.7 mg/mL glucose, 0.4 mg/mL glucose catalase, 0.08 mg/mL glucose oxidase). The final concentrations of K365 opto-kinesin was 450 nM. The final concentration of kinesin-14 was 125 nM. The sample was photoactivated using Lumencor Sola light engine through a 472 $\pm30$ nm excitation filter at intensity of 10 $\mu$W/mm$^2$. 

\subsection{Patterning active stress}
For bulk activation experiments, the epi-fluorescence arm of the microscope was used to illuminate the sample with blue light (Lumencor, Sola light engine white LED, 472 $\pm30$ nm excitation filter). Initially, the entire chamber was exposed to 37 mW/mm$^2$ of continuous blue light using a Lumencor Sola light engine through a $1\times$ 0.04 NA objective (Nikon Instruments, CFI Plan Achromat 1X)  for 10 minutes. This allowed the sample to obtain its isotropic bundled structure. Subsequent experiments were performed at $4\times$ 0.13 NA (Nikon Instruments, CFI Plan Fluor 4X) or $4\times$ 0.2 NA (Nikon Instruments, CFI Plan Apochromat Lambda D 4X) with pulsed blue light. 

For spatial patterning of activity, we used  a laser scanning confocal (Leica Microsystems, SP8) to the activate regions of arbitrary shape with a laser tuned to 488 nm (Leica Microsystems, Stellaris White Light Laser). Alternatively, an opaque mask was placed above the sample and a white LED (Nikon Instruments, Dia LED) was used to photoactivate the exposed region from above. The data in Fig. 5  was taken using opaque masks. For all experiments, the integrated intensity was calculated by first, measuring the power at the sample plane with a power meter. The entire sensor was filled by light, so that the intensity of light $I_{\textrm{raw}}=\frac{\textrm{power}}{\textrm{area of sensor}}$. To account for the pulsed activation, 
\begin{equation}
    I_{\textrm{integrated}} = I_{\textrm{raw}}\times \frac{\textrm{exposure time}}{\textrm{frame interval}} .
\end{equation}
This assumption is valid only within the regime where the motors are not unbinding on the timescales of the frame interval. We found this threshold to be 30 s from an experiment in which we incrementally increased the time between photoactivation pulses.  

\subsection{Analysis of Microtubule Flows}
Flow fields were obtained through particle image velocimetry (PIVLab MATLAB plugin) \cite{thielicke2014pivlab}. The PIV analysis parameters  were optimized for each data set [Fig.~S1]. Generally, a small window size in pixels resulted in larger average speeds. Thus, we kept the window size constant across data sets. By incrementally skipping frames in the analysis, we determined the optimal sampling of data to suppress noise and maintain temporal and spatial resolution. The average speed for each frame was calculated within the center of the chamber. 

For Fig. 2, the average speeds were obtained from imaging samples prepared as described above for 2-6 hours during which they were continuously photoactivated. The average was taken over the entire imaging time. 

\subsection{Analysis of Network Relaxation}
For Fig. 3, the average orientation of the network was found from the structure tensor of the fluorescent image of the microtubules using the FIJI Image J plugin OrientationJ with a window size of 8 pixels \cite{OrientationJ2016}. The initial angle was set to zero radians, defined to be along the $x$ axis. The standard deviation reported in Fig. 3(b) is from the spatially resolved orientation at a particular time.

\subsection{Calculation of strain fields}
The displacement and strain fields were calculated from the fluorescent images of the microtubules using NCORR2 in MATLAB \cite{NCORR2015}. The transition to instability was defined by the in-plane component of the strain perpendicular to the network alignment, $\gamma_{yy}$. Over the 17 minute activation for each intensity, we defined the onset of buckling as when the strain increase beyond 0.5\%. Below this threshold, the strain fluctuated around the noise floor [Fig.~S11]. 

\section{Relaxation of the bend instability upon deactivation}
We measured the average angle of a flow aligned microtubule network driven by non-processive opto-kinesin. Upon deactivation, the angle partially relaxed towards a more uniform state. We describe a model that accounts for the lingering activity of motors that are being deactivated over time and the elastic decay of the microtubule network. 

A function $M(t)$ describes the fraction of unbound motors where $t=0$ is the last moment of photoactivation. Thus, $M(t=0) = 0$ indicates that the bound motor fraction is completely saturated. In the dark, we will assume that $M(t\rightarrow\infty)=1$ indicating that the unbound motor fraction is completely saturated. $M(t)$ should take on a form that is consistent with the unbinding kinetics of proteins. We define
\begin{equation}
    M(t)=1-e^{-t/\tau} .
\end{equation}
Absorbance assay revealed the thermal reversion kinetics of the iLID domain is exponential decaying with characteristic time scale $\tau=24$ s assuming exponential kinetics [Fig.~\ref{S_iLID_unbinding_fit}]~\cite{Hallett2016} . It follows that the angle relaxation is described by 
\begin{equation}
    \langle\theta\rangle = (1-M(t))A(t) + R(t)
\end{equation}
\begin{equation}
    \langle\theta\rangle = e^{-t/\tau}A(t) + R(t)
\end{equation}
where $A(t)$ is the activity-driven growth of the angle and $R(t)$ is the elastic relaxation of the network. We measured $A(t)$ during the activation cycle and found $A(t)=\gamma t$, where $\gamma = 0.002$ rad/s [Fig.~\ref{S_AngularGrowth}]. 

We assume that the microtubule network behaves as a Hookean elastic solid such that its relaxation can be described by an overdamped spring 
\begin{equation}
    \ddot{x}+\beta\dot{x}+\omega_ox=0
\end{equation}
where the solution is given by the ideal limit conditions
\begin{equation}
R(t) =C*e^{-t/r}
\end{equation}
where $r=2\left(-\beta-\sqrt{\beta^2-4\omega_o^2}\right)^{-1}$ where $\beta$ describes the frictional dissipation and $\omega_o$ is the angular frequency of oscillation. 

Putting this together, we fitted the measured decay curves to:
\begin{equation}
    \langle\theta\rangle = e^{-t/\tau}\dot{\gamma} t+e^{-t/r}
\end{equation}
where $r$ is a fit parameter describing the characteristic time for the solid's relaxation. To compare relaxations at different stages of the bend instability we normalize the angle $\langle\theta_\textrm{norm}\rangle =\frac{ \theta_\textrm{i}-\theta_\textrm{min}}{\theta_\textrm{max}-\theta_\textrm{min}}$.

\section{Hydrodynamic Model of Confined Aligned Microtubule Network}
We consider a flow aligned active nematic with spatially patterned active stresses. In the illuminated region, the extensile active stresses destabilize orientational order. Let the instantaneous axis of orientational order of the microtubule bundles at position $\vec{r}$ be denoted $\hat{n}(\vec{r},t)$. $\hat{x}$ is the axis of initial nematic order, i.e., $\hat{n}(t=0)=\hat{x}$. Consider a spatially varying perturbation to the orientationally ordered state, such that $\hat{n}=\hat{x}+\delta \vec{n}_\perp(\vec{r})$.  Modelling the illuminated region using the hydrodynamic theory for an active nematic reveals that the eigenvectors of the dynamics correspond to Fourier transforms of twist-bend and splay-bend deformations, and that twist-bend modes grow faster than splay-bend modes \cite{Pooja2020}. Therefore, the instability seen in the experiment should correspond to a growing twist-bend deformation. Further, the dynamics of a twist-bend deformation with wavevector $\vec{k}$ is given by (see SI of Chandrakar et. al., 2020 \cite{Pooja2020} for the hydrodynamic model and the linear stability analysis)
\begin{align}
\partial_t[\hat{x}.(\vec{k}\times\delta \tilde{\vec{n}}_\perp)]=\left(-D_R\kappa  k^2+\frac{2\alpha}{3\eta}\frac{k_x^2}{k^2}\right) [\hat{x}.(\vec{k}\times\delta \tilde{\vec{n}}_\perp)]\label{growth_rate}
\end{align}
where $\alpha$ is the ``activity'' which corresponds to the force dipole generated by the extending microtubule bundles, $D_R$ is the rotational diffusion constant of the microtubule bundles, $\kappa$ is the nematic elasticity (assuming that bend, twist, and splay, all have the same energetic cost), and $\eta$ is the viscosity.
Let $H$ be the height of the chamber, $L$ be the size of the illuminated region along $\hat{x}$, and $W$ be the size of the illuminated region along $\hat{y}$. Then, the components of the wave vector are quantized by the confinement dimension, and are restricted to be such that $k_x>\pi/L, k_y>\pi/W,\ k_z>\pi/H$. 
Based on experimental observations, we assume $k_y=\pi/W$ and $k_z=\pi/H$. Then, depending on the value of $L$, there are two instability regimes: 
\begin{enumerate}
 \item At large $L$, the instability sets in as soon as the activity is high enough for the fastest growing wave mode from equation \ref{growth_rate} to be unstable. Taking the derivative of the growth rate in equation \ref{growth_rate} with respect to $k_x$ and equating it to zero, the fastest growing mode satisfies
 \begin{align}
\frac{\alpha(S_0+\xi)}{2\eta D_R\kappa}=\frac{(k_x^2+(\pi/W)^2+(\pi/H)^2)^2}{(\pi/W)^2+(\pi/H)^2}\label{fastest}
\end{align}
and is unstable (equation \ref{growth_rate} with this value of $k_x$ is positive) only for
\begin{align}
 \frac{\alpha(S_0+\xi)}{2\eta D_R\kappa}>4((\pi/W)^2+(\pi/H)^2) \label{condn}
\end{align}
From eqns \ref{fastest} and \ref{condn}, the fastest growing mode, if unstable has to satisfy
\begin{align}
k_x^2>(\pi/W)^2+(\pi/H)^2\label{cond2}
\end{align}
If $L$ is large enough that $(1/L)^2<(1/W)^2+(1/H^2)$, equation \ref{cond2} implies that $k_x>\pi/L$, so the fastest growing wave mode is not disallowed by confinement in the x dimension. Then, the threshold activity for the instability is given by equation \ref{condn}. 
To summarize, for large $L$ (i.e., $(1/L)^2<(1/W)^2+(1/H^2)$), the activity threshold for the instability is given by $\frac{\alpha(S_0+\xi)}{2\eta D_R\kappa}>4\pi^2\left(\frac{1}{W^2}+\frac{1}{H^2}\right)$

 \item For small $L$ ($\frac{1}{L^2}>\frac{1}{W^2}+\frac{1}{H^2}$), even when the activity is high enough for the emergent fastest growing mode given by equation \ref{fastest} to be unstable, there is no instability because this emergent mode is disallowed by the boundaries. Instead, the instability is observed when the growth rate of the $k_x=\frac{\pi}{L}$ mode becomes positive, i.e. (from eq. \ref{growth_rate}),
 \begin{align}
  \frac{\alpha(S_0+\xi)}{2\eta D_R\kappa}>\pi^2L^2\left(\frac{1}{L^2}+\frac{1}{W^2}+\frac{1}{H^2}\right)^2
 \end{align}
\end{enumerate}

\newpage

\section{Supplementary Figures}


\noindent\textbf{Supplementary Fig. 1: PIV flow fields from fluorescence images.} 

\noindent\textbf{Supplementary Fig. 2: Reduced difference between on and off flows in active gels driven by processive opto-kinesin upon repeated photoactivation.} 

\noindent\textbf{Supplementary Fig. 3: Kinesin motor coding regions.}

\noindent\textbf{Supplementary Fig. 4: In the dark, the measured microtubule flows for the K365 opto kinesin are zero.}

\noindent\textbf{Supplementary Fig. 5: Measuring network angle from microtubule fluorescence using OrientationJ.}

\noindent\textbf{Supplementary Fig. 6: Measuring the angular growth rate from the onset of the bend instability.}

\noindent\textbf{Supplementary Fig 7: Extracting reversion timescale from absorbance recovery of iLID construct.}

\noindent\textbf{Supplementary Fig. 8: Fitting the microtubule network decay to the elastic model.}

\noindent\textbf{Supplementary Fig. 9: Displacement and strain fields of spatially patterned active stress.}

\noindent\textbf{Supplementary Fig. 10: Depletion induced contraction of a flow aligned sample.} 

\noindent\textbf{Supplementary Fig. 11: Defining threshold intensity for instability.}

\noindent\textbf{Supplementary Fig. 12: Flow fields after de-activation are highly correlated with pre-activation flows.}

\noindent\textbf{Supplementary Fig. 13: Hysteresis in intensity-controlled active flow speed.}

\newpage
\begin{figure}
    \centering
    \includegraphics[width=12cm]{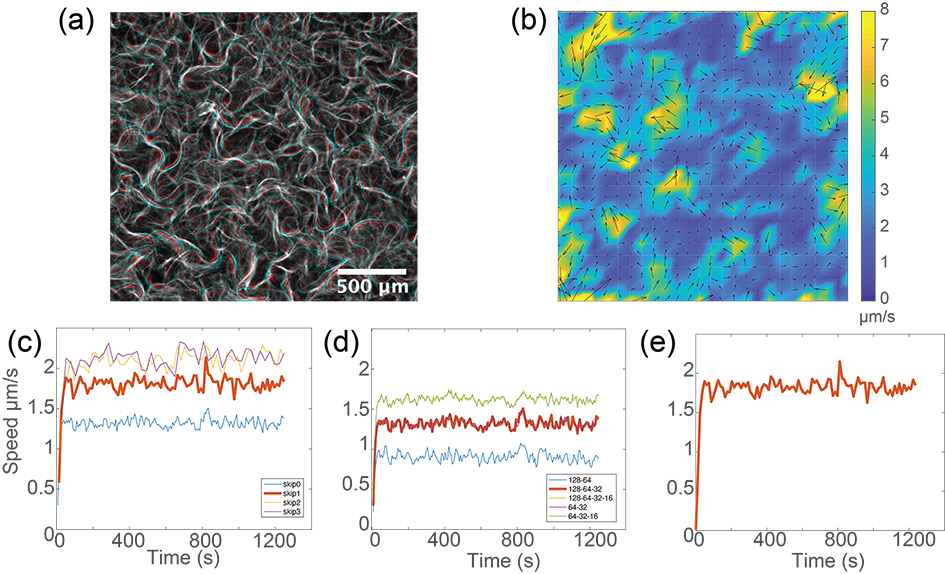}
    \caption{\textbf{PIV flow fields from fluorescent microtubule images.} \textbf{(a)} Overlay of two images taken at t=0 (red) and t=5 s (cyan) illustrate the dynamical microtubule network. \textbf{(b)} Flow field of images in (a) where colormap shows speed. \textbf{(c)} Average speed versus time for various PIV window sizes. \textbf{(d)} Average speed versus increasing frame interval. Parameters used for this data set are represented by bolded curve. \textbf{(e)} The chosen optimal mean speed versus time curve for this data set. }
    \label{S_PIV}
\end{figure}
\begin{figure}
    \centering
    \includegraphics[width=9cm]{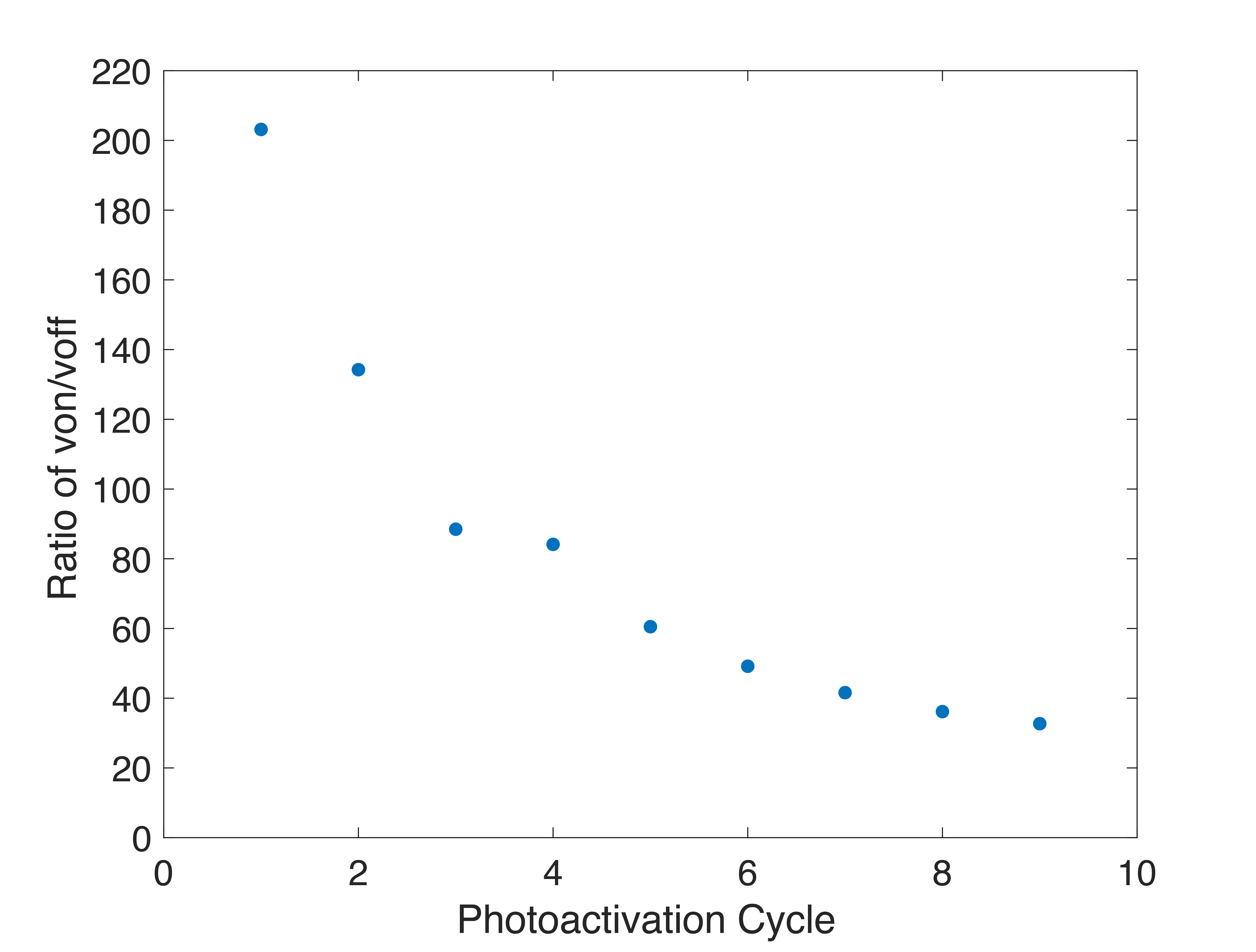}
    \caption{\textbf{Aging of active gels powered with K401 opto-kinesin clusters.} The ratio of microtubule flow speeds in the illuminated and dark state as a function of the photoactivation cycle.}
    \label{S_von-voff_ratio}
\end{figure}
\begin{figure}
    \centering
    \includegraphics[width=8cm]{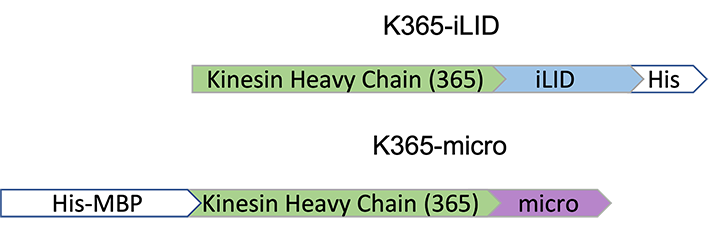}
    \caption{\textbf{Kinesin motor coding regions } }
    \label{S_genemap}
\end{figure}
\begin{figure}
    \centering
    \includegraphics[width=6cm]{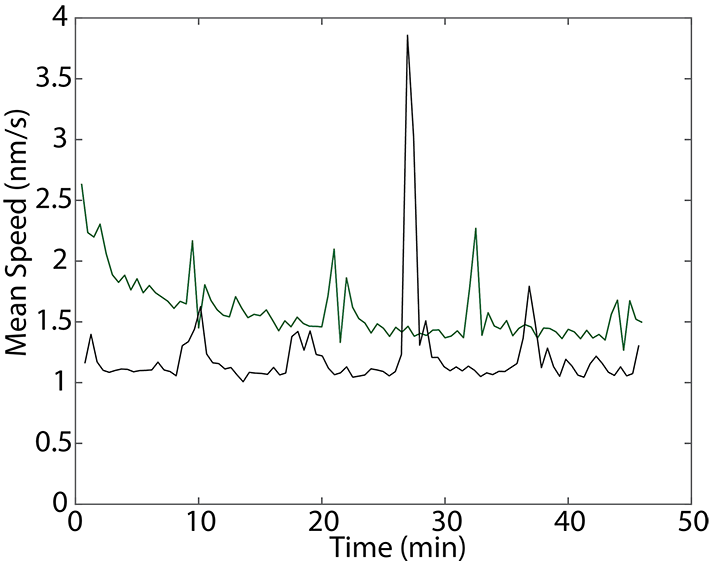}
    \caption{\textbf{Dark-state dynamics of K365 opto-kinesin active fluids.} The velocity of active fluid where the sample is in the dark (green) and a sample that ran out of ATP (black). Both analyses were done at 30 s/frame and the same PIV settings.}
    \label{S_DarkSpeed}
\end{figure}
\begin{figure}
    \centering
    \includegraphics[width=12cm]{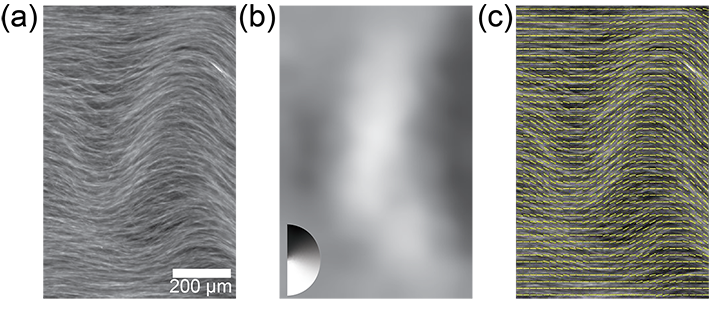}
    \caption{\textbf{Measuring network angle from microtubule fluorescence using OrientationJ.} (a) Fluorescence image of shear-aligned microtubules undergoing bend instability. (b) Filament orientation obtained from the structure tensor analysis using OrientationJ with window size of 8 pixels. Intensity indicates angle. (c) Calculated orientation field plotted over fluorescence image.}
    \label{S_OrientationJ}
\end{figure}
\begin{figure}
    \centering
    \includegraphics[width=8cm]{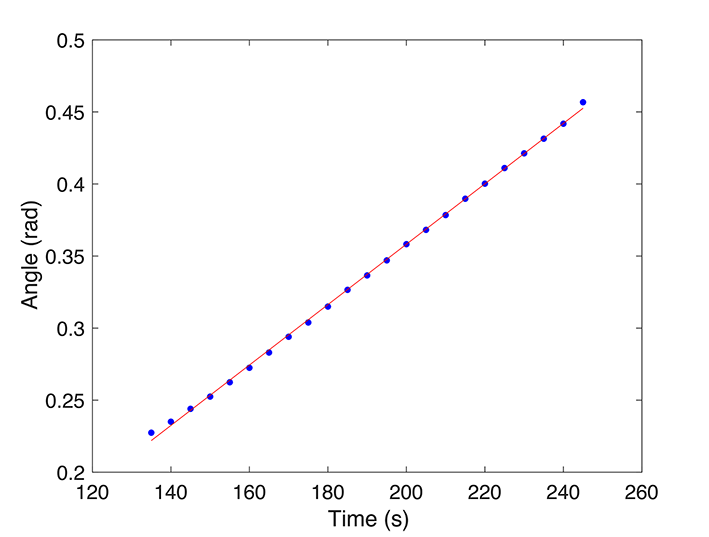}
    \caption{\textbf{Measuring angular growth rate from the onset of the bend instability.} The filament orientation angle versus time during activation of the bend instability. The line is a fit with the slope $\gamma = 0.002$ rad/s. }
    \label{S_AngularGrowth}
\end{figure}

\begin{figure}
    \centering
    \includegraphics[width=6cm]{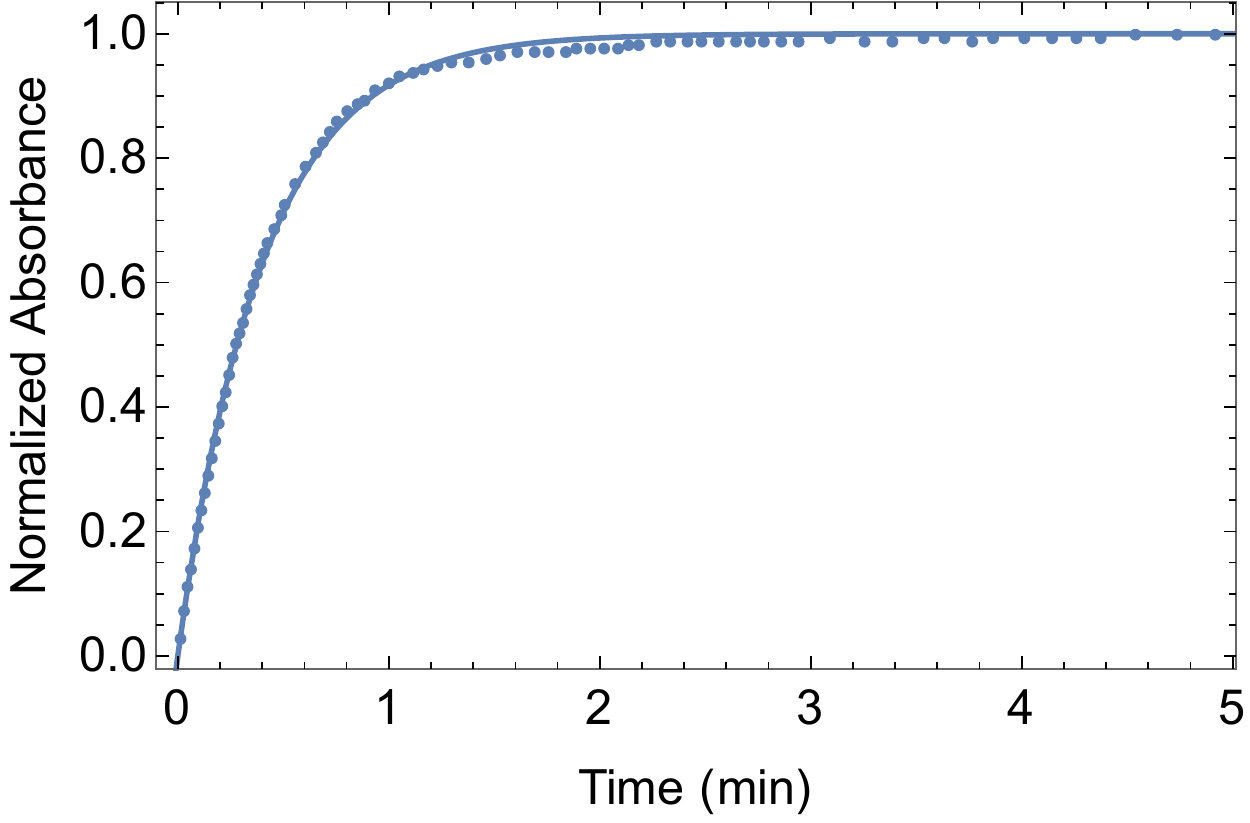}
    \caption{\textbf{Extracting reversion timescale from absorbance recovery of iLID construct.} Modified from \cite{Hallett2016}. A plot of the absorbance at 450 nm recovery after activation for the iLID construct where an absorbance of 1 indicates a full recovery to the dark structure of the iLID protein. The line indicates a fit to a bounded exponential $A=1-e^{-t/\tau}$ where $\tau = 24$ s. }
    \label{S_iLID_unbinding_fit}
\end{figure}
\begin{figure}
    \centering
    \includegraphics[width=12cm]{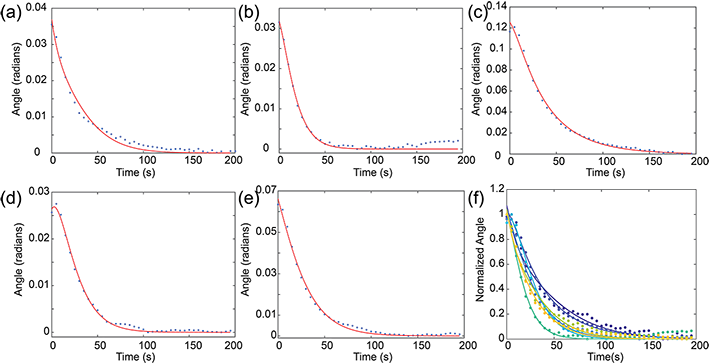}
    \caption{\textbf{Fitting the microtubule network decay to the elastic model.} (a-e) After de-activation a sample undergoing bend-instability undergoes partial relaxation. Fits of average angle $\langle\theta\rangle$ to Equation 1 after de-activation. (f) Data and fits were normalized by the maximum angle such that $\langle\theta_{\textrm{norm}}\rangle = \frac{\langle \theta \rangle - \theta_{\textrm{min}}}{\theta_{\textrm{max}}-\theta_{\textrm{min}}}$. The average is shown in Fig. 3(b).}
    \label{S_ElasticFit}
\end{figure}
\begin{figure}
    \centering
    \includegraphics[width=12cm]{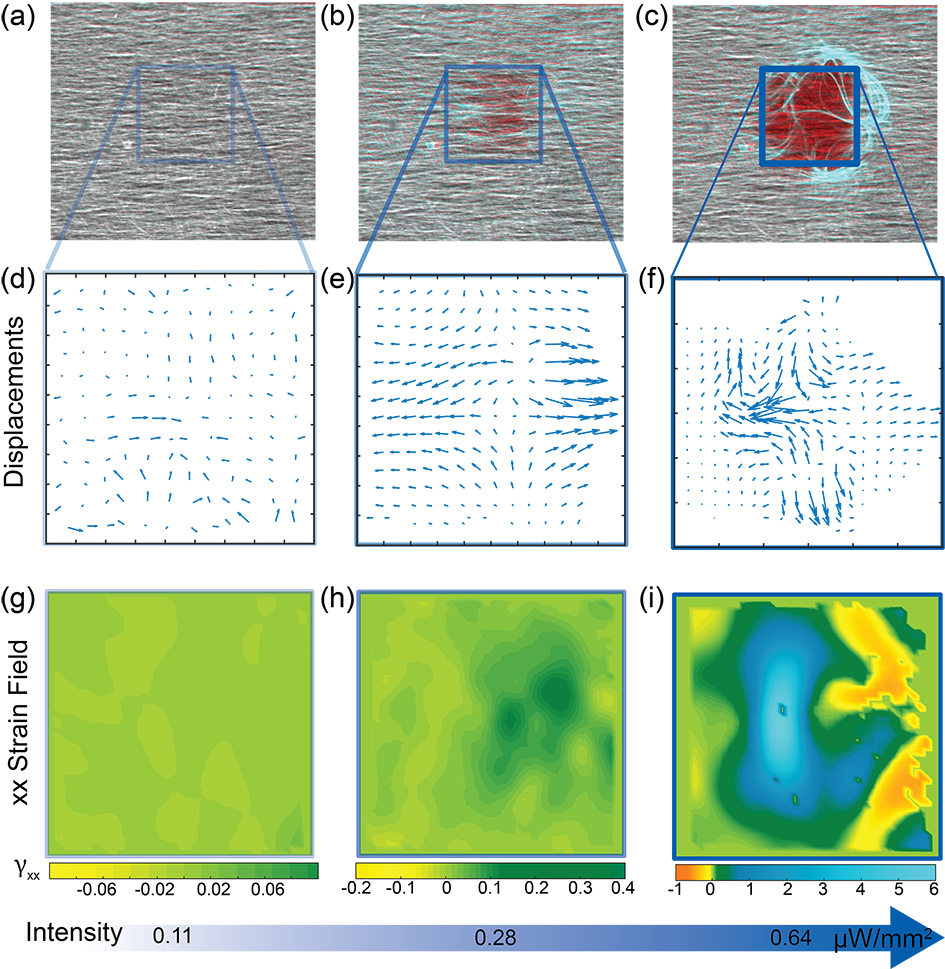}
    \caption{\textbf{Displacement and strain fields of spatially patterned active stress.} \textbf{(a-c)} Overlay of initial (red) and final (cyan) of a flow aligned 3-dimensional network of microtubules where the highlighted region is activated with light. Gray indicates no movement of filaments (red + cyan = white). Indicated squares are 400 $\times$ 400 $\mu$m$^2$. \textbf{(a)} Quiescent regime: the intensity in the activated region, indicated by the yellow square, is low 0.11$\mu$W/mm$^2$. \textbf{(b)} Sliding regime: the intensity in the activated region, indicated by the green square, is 0.28 $\mu$W/mm$^2$. \textbf{(c)} Turbulent regime: the intensity in the activated region, indicated by the blue square is high 0.64 $\mu$W/mm$^2$. \textbf{(d-f)} Displacement field of the activated regions in (a-c) respectively. \textbf{(g-i)} Strain field $\gamma_{xx}$ measured from the displacements in (d-f) respectively. }
    \label{S_strain}
\end{figure}
\begin{figure}
    \centering
    \includegraphics[width=12cm]{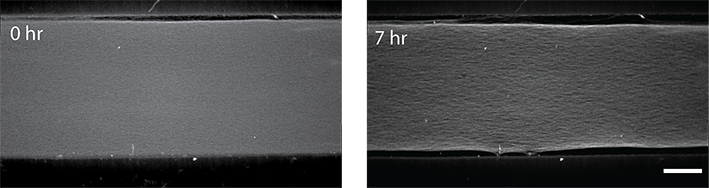}
    \caption{\textbf{Depletion induced contraction of a flow aligned sample.} Snapshots of an aligned microtubule network exhibiting depletion induced contraction.}
    \label{S_Depletion}
\end{figure}

\begin{figure}
    \centering
    \includegraphics[width=8cm]{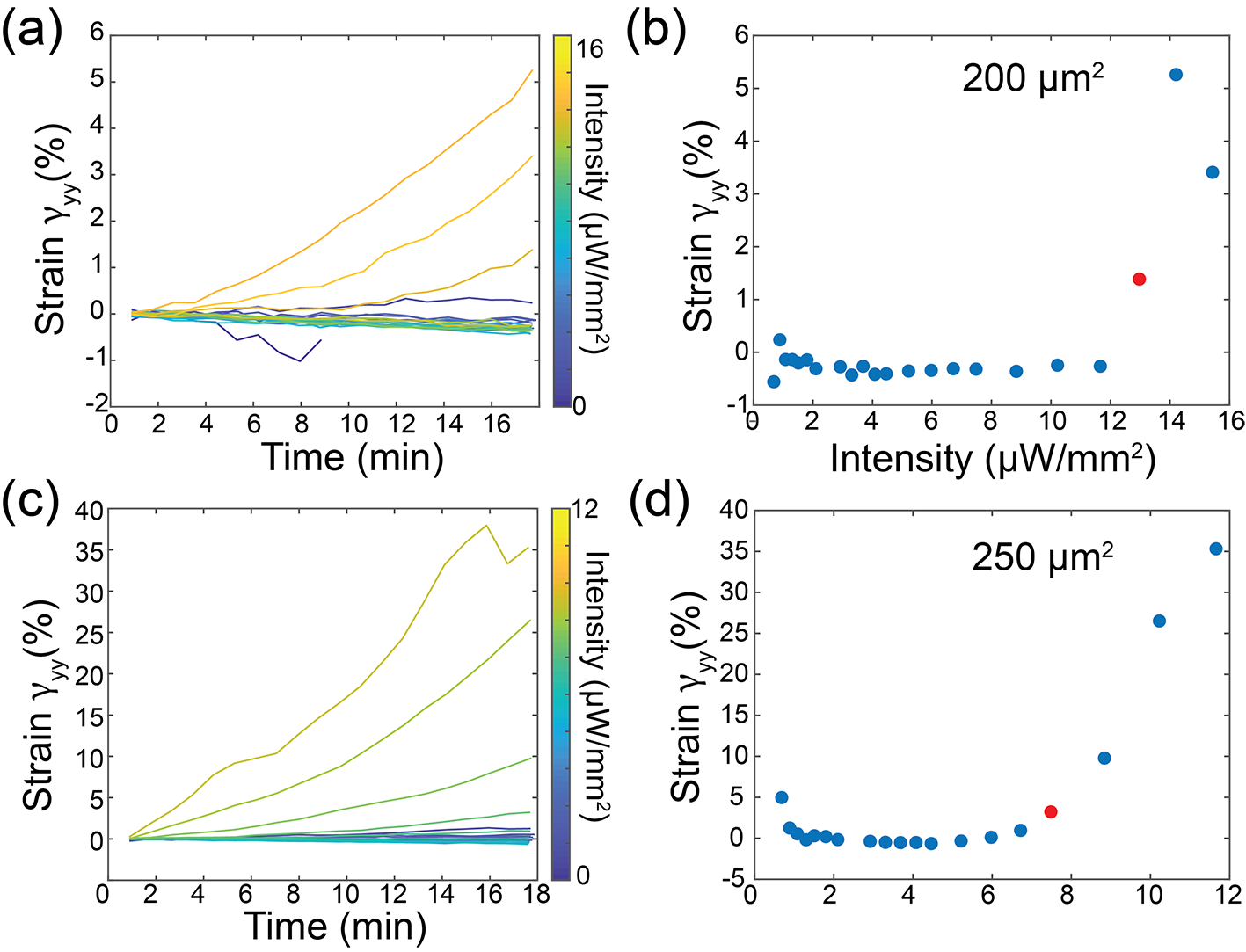}
    \caption{\textbf{Defining threshold intensity for instability.} \textbf{(a)} The component of strain perpendicular to alignment $\gamma_{yy}$ as function of time for a 200 $\mu$m$^2$ activated region. Color indicates the intensity of activation light at 488 nm. \textbf{b} The strain $\gamma_{yy}$ accumulated after 17 min of activation as a function of activation intensity for a 200 $\mu$m$^2$ activated area. The red data point indicates the intensity threshold for instability. \textbf{(c-d)} The same measurements for a 300 $\mu$m$^2$ activation region. The red data point indicates the threshold intensity for the instability. }
    \label{S_Thresh}
\end{figure}

\begin{figure}
    \centering
    \includegraphics[width=12cm]{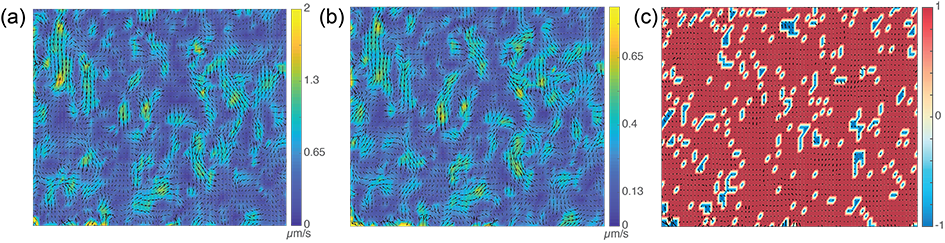}
    \caption{\textbf{Flow fields after de-activation are highly correlated with pre-activation flows.} (a) Flow field in last frame of of a light cycle PIV. Color indicates flow speed. (b) Re-activated flow field after 30 min dark cycle. (c) The difference between the two velocities. The correlation coefficient between the two flow fields is 0.89. }
    \label{S_FlowFieldCorr}
\end{figure}

\begin{figure}
    \centering
    \includegraphics[width=12cm]{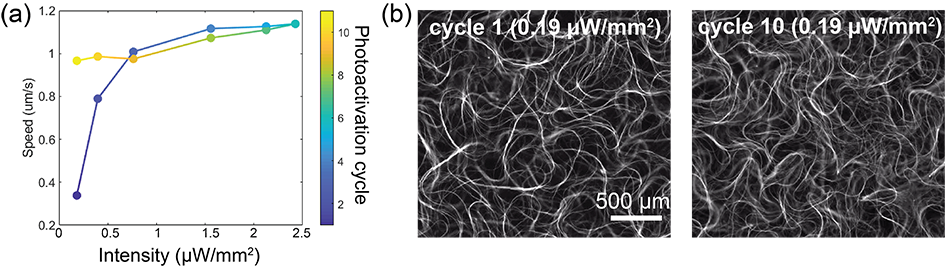}
    \caption{\textbf{Hysteresis in intensity-controlled active flow speed.} \textbf{(a)} The flow speed plotted as a function of photoactivation intensity. The color bar indicates forward in time from blue to yellow. \textbf{(b)} Bundle structure at the same photoactivation intensity for the cycle 1 and cycle 10 respectively.}
    \label{S_Hysteresis}
\end{figure}

\clearpage
\newpage

\section{Video Captions}

Videos are located \href{https://drive.google.com/drive/folders/1PxWHA9pl1gi3Ur0zZ28VEByDGYBMmPUY?usp=sharing}{here}.

\textbf{Video 1: K401 opto-kinesin drive extensile active fluid dynamics when photoactivated.} This video shows fluorescently labeled microtubules in an active fluid alternately photoactivated by 2.4 $\mu$W/mm$^2$ blue light and deactivated. Scale bar: 500 $\mu$m.  


\textbf{Video 2: K365 opto-kinesin drive extensile active fluid dynamics.} This video shows fluorescently labeled microtubules in an active fluid alternately photoactivated by 2.2 $\mu$W/mm$^2$ blue light and deactivated. Scale bar: 500 $\mu$m. 


\textbf{Video 3: Controlling the onset of the bend instability.} An initially aligned microtubule network with 300 nM K365 opto-kinesin was exposed to bursts of photoactivation which pushed the microtubules into the bend instability. Subsequent deactivation showed relaxation of the passive microtubule network as the K365 opto-kinesin un-clustered. 

\textbf{Video 4: Size dependence of bend instability.} An initially aligned network of fluorescently labeled microtubules and K365 opto-kinesin was photoactivated inside the yellow boxes of various sizes (50 $\mu$m$^2$ - 500 $\mu$m$^2$). At a single photoactivation intensity, the largest box (500 $\mu$m$^2$) underwent the bend instability, the mid-size boxes (250 $\mu$m$^2$, 200 $\mu$m$^2$) mostly exhibited sliding dynamics and the smaller boxes (100 $\mu$m$^2$ and 50 $\mu$m$^2$) remained quiescent. 

\textbf{Video 5: Confined flows in aligned microtubule network for quiescent, sliding and bend instability regimes.} An initially aligned network of fluorescently labeled microtubules was sequentially photoactivated at increasing intensities within a 400 $\mu$m$^2$ box indicated by the yellow border. Left panel: quiescent regime at low activation intensity, 0.11 $\mu$W/mm$^2$. Middle panel: sliding regime at 0.28 $\mu$W/mm$^2$. Right panel: bend instability regime at high activation intensity, 0.64 $\mu$W/mm$^2$. 

\textbf{Video 6: Depletion of microtubules within activated region.} An initially aligned network of fluorescently labeled microtubules and K365 opto-kinesin with a photobleached line as a material marker. The region inside the yellow box was photoactivated with a 488 nm laser. The microtubules undergo successive bend instabilities and eventually deplete from the activated region. 

\textbf{Video 7: Kinesin-14 and K365 opto-kinesin without activation contracts for many hours.} An active network of fluorescently labeled microtubules with both kinesin-14 and K365 opto-kinesin motors. The sample was never photoactivated so that the K365 opto-kinesin were un-clustered. The network contracts for the acquisition time, 12 hours. 

\textbf{Video 8: Transition from contractile to extensile active stress in composite motor system.} An active network of fluorescently labeled microtubules with kinesin-14 motors and K365 opto-kinesin motors. Initially the K365 opto-kinesin were not activated, so that the network contracted into a dense aligned bundle driven by kinesin-14 motors. Upon photoactivation, the material extended and underwent the bend instability. 

\textbf{Video 9: Extended illumination below ``threshold'' intensity within confined region eventually leads to the bend instability.} An initially aligned microtubule network is exposed to low intensity photoactivation light within the indicated 400 $\mu$m$^2$ square. Eventually, the microtubules slide out of the activated region and the material bends perpendicular to the direction of alignment.

\clearpage
\newpage

\end{document}